# POTENTIAL VORTICITY OF SATURN'S POLAR REGIONS: SEASONALITY AND INSTABILITIES

Arrate Antuñano[a], Teresa del Río-Gaztelurrutia[b], Agustín Sánchez-Lavega[b], Peter L. Read[c], Leigh N. Fletcher[a]

[a] Department of Physics & Astronomy, University of Leicester, University Road, Leicester LE1 7RH, UK.

[b] Departamento de Física Aplicada I, Escuela de Ingeniería de Bilbao, Universidad del País Vasco, Bilbao, Spain.

[c] Clarendon Laboratory, University of Oxford, Parks Road, Oxford, OX1 3PU, UK.

**Key Points**

- Quasi-geostrophic potential vorticity maps of Saturn's polar regions are presented.
- Meridional gradients of quasi-geostrophic potential vorticity cannot explain the existence of the Hexagon only in the northern hemisphere.
- Quasi-geostrophic potential vorticity maps do not show significant seasonal variability.

**Abstract**

We analyse the potential vorticity of Saturn's polar regions, as it is a fundamental dynamical tracer that enables us to improve our understanding of the dynamics of these regions and their seasonal variability. In particular, we present zonally averaged quasi-geostrophic potential vorticity maps between 68° planetographic latitude and the poles at altitudes between 500 mbar and 1mbar for three different epochs: (i) June 2013 (early northern summer) for the north polar region, (ii) December 2008 (late northern winter) for both polar regions and (iii) October 2006 (southern summer) for the south, computed using temperature profiles retrieved from Cassini Composite Infrared Spectrometer (CIRS) data and wind profiles obtained from Cassini's Imaging Science Subsystem (ISS). The results show that quasi-geostrophic potential vorticity maps are very similar at all the studied epochs, showing positive vorticities at the north and negative at the south, indicative of the dominance of the Coriolis parameter $2\Omega\sin\phi$ at all latitudes, except near the pole. The meridional gradients of the quasi-geostrophic potential vorticity show that dynamical instabilities, mainly due to the barotropic term, could develop at the flanks of the Hexagon at 78°N, the jet at 73.9°S and on the equatorward flank of both polar jets. There are no differences in potential vorticity gradients between the two hemispheres that could explain why a hexagon forms in the north and not in the south. No seasonal variability of the potential vorticity and its meridional gradient has been found, despite significant changes in the atmospheric temperatures over time.

1. **Introduction**

Saturn's rotation axis is tilted at an angle of 26.7° relative to its orbital plane, causing substantial seasonal variations of insolation over a 29.5-year cycle (one Saturnian solar orbit). Saturn's polar regions are limited by the latitude circles of ~68° planetographic latitude north and south. Within these polar regions, the dynamics at the upper cloud level is dominated by a permanent narrow eastward jet present in both hemispheres at 78°N and 73.9°S, and by an intense cyclonic circulation centered at each pole (*Antuñano et al., 2015; Sayanagi et al., 2017*). Furthermore, Saturn's polar regions present a large variety of cloud morphologies, including the hexagonal wave in the north polar region at 75.8° planetocentric latitude (*Sánchez-Lavega et al., 2014; Antuñano et al., 2015; Sayanagi et al., 2016*) and two stable polar cyclones, one at each pole (*Sánchez-Lavega et al., 2006; Dyudina et al., 2008; Fletcher et al., 2008; Baines et al., 2009; Antuñano et al., 2015; Sayanagi et al., 2017*).

Although some transient local cloud activity has been observed at the Hexagon jet and the north polar vortex (*Antuñano et al., 2018*), the zonal wind profiles of the polar regions have remained unchanged over the last 35 years, without any signs of seasonal variations (*Sánchez-Lavega et al., 2000; García-Melendo et al., 2011; Antuñano et al., 2015; Sayanagi et al., 2016; Sayanagi et al., 2017*). On the other hand, observations of Saturn's atmosphere in the mid and far infrared (7-1000 μm) over the last 35 years, have shown a large seasonal asymmetry in the polar temperature field between summer and winter at all altitudes above the 500-mbar pressure level (*Conrath and Pirraglia, 1983; Bézard et al., 1984; Yamanandra-Fisher et al., 2001; Fletcher et al., 2016*), with the presence of a warm stratospheric polar hood at the summer pole (*Orton and Yanamandra-Fisher, 2005; Fletcher et al., 2007*). Using the thermal wind equations (assuming geostrophy), previous analysis of the temperature and wind profiles of the polar regions have shown that the vertical wind-shear is negative (i.e., increasingly retrograde) at eastward jets and positive (i.e., increasingly prograde) at westward jets, reducing the magnitudes of the jets with altitude, regardless of their direction (*Greathouse et al., 2005; Read et al., 2009; Fletcher et al., 2016*). A long-term study of the thermal wind variability at 100 mbar over a third of Saturn's year by *Fletcher et al.* (*2016*) showed that the windshear was largely unchanged during southern summer and autumn, whilst larger changes were observed in the northern hemisphere as it moved from winter into spring, particularly at the retrograde jets.

Potential vorticity (e.g. *Vallis, 2006*), in its various definitions, is a fundamental quantity to characterize the dynamical state of an atmosphere, as it is conserved under different approximations, with Ertel potential vorticity (EPV) conserved in adiabatic motion of an atmosphere of unchanging composition, and quasi-geostrophic potential vorticity (QGPV) conserved in geostrophic horizontal motion (*Vallis, 2006*). This implies that the potential vorticity (PV) can be used as a tracer of atmospheric motions. Moreover, with adequate boundary conditions, the state of the atmosphere (including winds and thermal structure) can, in principle, be retrieved from a knowledge of the PV field (*Sánchez-Lavega, 2011*). Finally, the structure of the PV field gives hints on potential instabilities of the atmosphere, which can provide clues about the form of the deep flow in the giant planets.

*Read et al.* (*2009*) presented the first zonally averaged Ertel and quasi-geostrophic potential vorticity maps of Saturn's upper troposphere and stratosphere (~ 500-1 mbar) between 90°S and 78°N, using the zonally averaged temperature profiles retrieved from Cassini CIRS (*Fletcher et al., 2007*) and the horizontal wind field of the north hemisphere from the Voyager flybys (*Godfrey, 1988; Sánchez-Lavega et al., 2000*) and of the south hemisphere from HST and Cassini (*Sánchez-Lavega, 2002; Vasavada et al., 2006*). In their work, they showed that both EPV and QGPV exhibit a step-like variation with latitude in horizontal pressure surfaces, associated with the jets, being most prominent in the troposphere. Furthermore, they confirmed that there were several latitudinal regions where the Rayleigh-Kuo (*Kuo, 1949*) and the Charney-Stern criterion (*Charney and Stern, 1962*) for barotropic/baroclinic instabilities to grow was satisfied, even when baroclinic effects, due to changes in the vertical gradient of the absolute vorticity, were taken into account. They also analysed the Arnol'd II criterion for instability *(Arnol'd, 1966)*, showing that most of the atmosphere of Saturn was close to marginal instability under this criterion, although there were latitudes, near locations where meridional gradients of the potential vorticity changed sign, where neutrality appeared to be violated.

Finally, *Fletcher et al.* (*2016*) analysed the seasonal evolution of the quasi-geostrophic potential vorticity gradients over a third of the Saturnian year (2004-2014) between 80°N and 80°S, showing that the potential vorticity gradients change significantly at mid latitudes, while no change was observed at equatorial latitudes. However, it is important to note that these calculations rely on gradients of the temperature field, both in the horizontal and the vertical and thus, they are often subject to rather large uncertainties.

In this study, we compute zonally averaged quasi-geostrophic potential vorticity maps of both Saturn's Polar Regions at latitudes higher than 68º planetographic and heights between the ammonia cloud tops (~500 mbar, *Sanz-Requena et al, 2017)*, where winds have been measured, and 1 mbar pressure levels. In our maps, we combine wind profiles measured from Cassini ISS by cloud tracking (*Sánchez-Lavega et al., 2006; García-Melendo et al., 2011; Antuñano et al., 2015*), with high resolution zonally averaged temperature profiles retrieved from Cassini CIRS data (*Fletcher et al., 2016*). At both poles, we compare two different epochs, late winter and early summer in the case of the north, and early and late summer in the case of the south. By computing the meridional gradient of the retrieved potential vorticities, we analyse the stability of the jets present at the Polar Regions above cloud tops. Finally, we analyse the temporal changes in the potential vorticity maps looking for possible seasonal effects, and we find that changes are mainly related to the seasonal evolution of the temperature fields, since zonal winds remain essentially stable at cloud level. This way, we provide a diagnostic of the behaviour of the polar dynamics in a planetary atmosphere different to Earth, but also subject to seasonal insolation changes, giving important information about the nature of waves and dynamical instabilities in those regions.

The current study extends the work of *Read et al*. (2009) to map the whole North Polar Region and includes information about the potential vorticity at different Saturnian epochs, essential for the understanding of seasonal effects at the three-dimensional dynamics at the Polar Regions. Moreover, we use higher resolution thermal and wind profiles, which might help the understanding of the very particular dynamics of Saturn's Polar Regions.

## 2. Database and Methodology
### 2.1. Data selection

In this study, we use high resolution wind measurements at cloud level obtained from images captured by the Imaging Science System (ISS) instrument (*Porco et al., 2004*) on-board the Cassini spacecraft, and zonally averaged temperature profiles retrieved from the Composite Infrared Spectrometer (CIRS) (*Flasar et al., 2004*). In particular, we use four different sets of wind profiles at cloud level (i) June 2013 (early northern summer) from latitudes 68° to 90° (all latitudes in this study are planetographic, *Antuñano et al., 2015*), (ii) December 2008 (late northern winter) from latitudes 70° to 84° (*García-Melendo et al., 2011*), (iii) October/December 2008 from latitudes -68° to -90° (*Antuñano et al., 2015*) and (iv) October 2006 (southern summer) from latitudes -68° to -90° (*Sánchez-Lavega et al., 2006*); and four datasets of zonally averaged temperature profiles for pressures between 1 mbar and 1 bar. Due to the lack of simultaneous measurements of winds and temperatures during the Cassini mission, the temperature fields for the months of interest (i.e. June 2013, December 2008 and October 2006) were obtained by interpolating the existing temperature measurements over the whole Cassini mission, and extracting only the months where wind measurements were available (see *Fletcher et al. 2015, 2016*). Temperatures were retrieved from CIRS focal plane 1 (10-600 cm$^{-1}$), sensing the 80-800 mbar range, and focal plane 4 (1100-1400 cm$^{-1}$), sensing the 0.5-5.0 mbar range. The maximum spatial resolution of the wind measurements is overall smaller than 0.5° latitude, while the maximum spatial resolution of the zonally averaged temperature profiles is approximately 2° latitude. The vertical resolution of the temperature profiles, which were derived from nadir spectroscopy, is approximately a scale height (*Fletcher et al., 2016*), implying that small-scale vertical variability (e.g., associated with waves) would be smoothed by the retrieval process, making the zonally averaged axisymmetric approach relevant.

Table 1. Dataset used in this study.

| Zonal wind profiles | | |
|---|---|---|
| **Date** | **Hemisphere** | **Reference** |
| June 2013 | North | Antuñano et al., 2015 |
| December 2008 | North | García-Melendo et al., 2011 |
| October 2006 | South | Sánchez-Lavega et al., 2006 |
| December 2008 | South | Antuñano et al., 2015 |

| Temperature profiles | | |
|---|---|---|
| **Date** | **Hemisphere** | **Reference** |
| June 2013 | North | Fletcher et al., 2016 |
| December 2008 | North | Fletcher et al., 2016 |
| October 2006 | South | Fletcher et al., 2016 |
| December 2008 | South | Fletcher et al., 2016 |

Figure 1 displays the zonally averaged temperature profiles used in this study. These temperature maps differ at each polar region mainly in the upper troposphere and stratosphere, while no significant seasonal effects are observed below 150 mbar. The temperature difference between the studied epochs is larger in the north polar region, with a maximum temperature difference of ~ 20 K in the stratosphere between 1 mbar and 8 mbar between December 2008 and June 2013, while an ~ 8 K difference is found in the south between October 2006 and December 2008. The temperatures from December 2008 (north) and October 2006 (south) used in this study are similar to those found in *Read et al.* (*2009*), although from different dates. However, the June 2013 dataset shows a ~ 20 K warming in the north polar stratosphere compared to the dataset used by *Read et al.* (*2009*), and the December 2008 dataset shows a ~ 8 K cooling in the southern hemisphere. Both of these trends are consistent with seasonal warming/cooling.

As the temperature data used in this work was not obtained on a regular grid, all maps presented in this study are built by creating a regular grid of 21x21 points and interpolating the data using a Delaunay triangulation in 1.5° latitude x 0.14 log(P/P$_0$) and then, evaluating the data in this new grid.

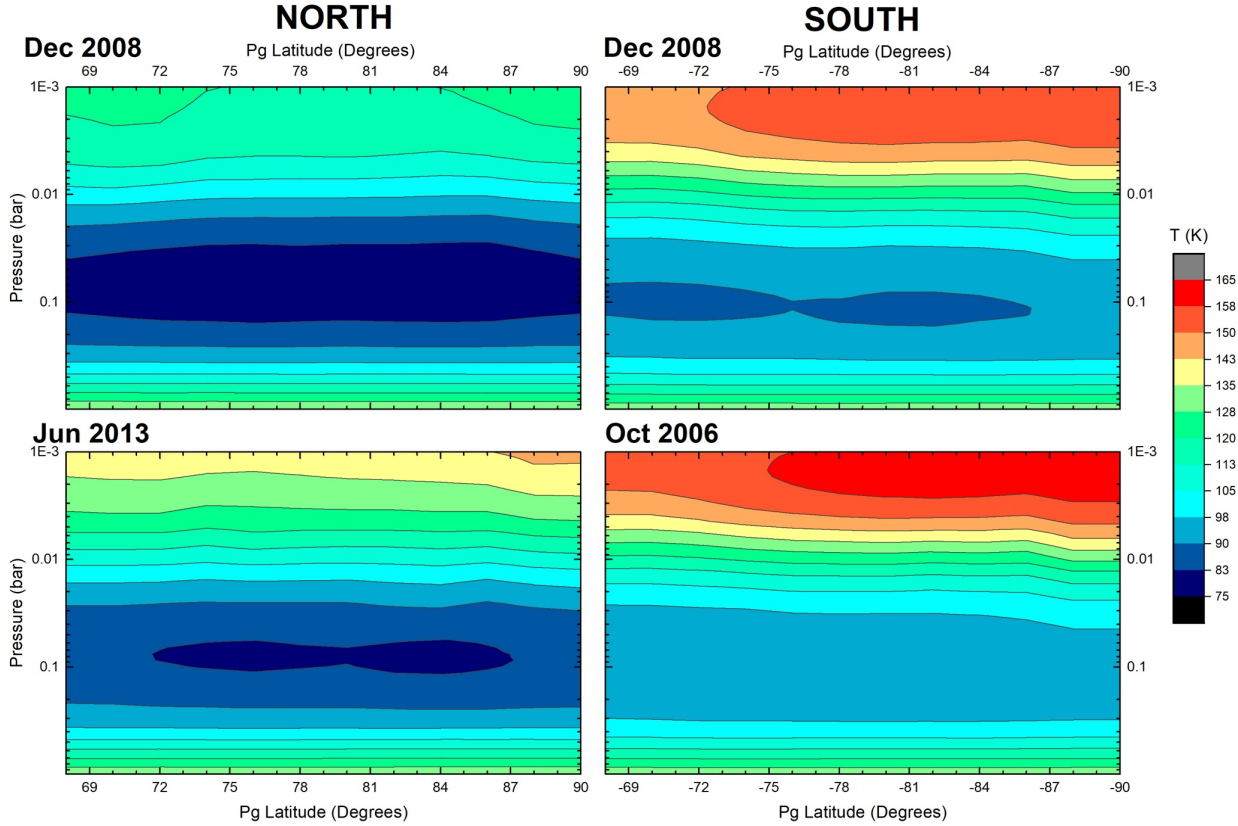

Figure 1. Zonally averaged temperature maps from 68° to 90° of the North Polar Region (left) and the South Polar Region (right) from December 2008 (top), June 2013 (left bottom) and October 2006 (right bottom).

## 2.2. Ortho-para fraction dependent specific heat capacity and definition of potential temperature

Usually, in adiabatic and frictionless motion, Ertel potential vorticity is conserved on isentropic surfaces. However, this conservation applies only to flows where the potential temperature depends only on pressure and density (e.g. *Gierasch et al., 2004*). When assuming perfect gas law behaviour and a constant specific heat capacity, the potential temperature has the simple analytical expression given by

$$\theta = T \left(\frac{P_0}{P}\right)^{\frac{R^*}{c_p^0}} \quad (1)$$

with $R^* = 3.885$ J g$^{-1}$ K$^{-1}$ being the specific gas constant on Saturn *(Sánchez-Lavega, 2011)* and $C_p^0$ the constant pressure specific heat capacity. Nevertheless, in the range of temperatures of the outer planets' atmospheres between 1 bar and 1 mbar, the specific heat capacity depends on the temperature and thus, equation (1) is no longer valid. Moreover, the hydrogen ortho-para fraction may vary, which leads to more complicated thermodynamics described by *Gierasch et al.* (*2004*). Here we will ignore ortho-para exchange and consider a frozen "normal" specific heat capacity in Saturn's atmosphere (*Sánchez-Lavega, 2011*), corresponding to a meta-stable distribution of hydrogen-helium with no ortho-para exchange and an assumed fraction of molecular hydrogen, H$_2$, populating the odd state of $X_{ortho} = 0.75$ (*Conrath and Gierasch, 1984*).

In order to obtain a potential temperature that depends only on pressure and density, we approximate the specific heat capacity of this frozen normal distribution by a power series by adjusting the coefficients empirically over the range of temperatures of interest. In the case of Saturn's atmosphere at the Polar Regions, we have approximated the specific heat capacity between 80 K and 160 K by a third order polynomial:

$$C_p/R^* \sim A + B\,T + C\,T^2 + D\,T^3 \qquad (2)$$

where A = 2.517755, B = -0.006530 K$^{-1}$, C = 1.195531 x 10$^{-4}$ K$^{-2}$ and D = -3.504449 x10$^{-7}$ K$^{-3}$.

This way, we arrive at an analytical expression for the potential temperature in the range of interest. In order to do so, we introduce a new magnitude $\tau$, with physical dimensions of temperature, defined as

$$C_p(T)\frac{\delta T}{T} = C_p^0 \frac{\delta \tau}{\tau} \qquad (3)$$

(*Epele et al., 2007, Garate-Lopez et al., 2015*), where $C_p^0 = C_p(T_0)$ and $T_0 = \tau(T_0)$. The relation between T and $\tau$ is obtained by replacing $C_p(T)$ with the approximation given in eq. 2 and then integrating eq. 3

$$\tau = T_0 \left(\frac{T}{T_0}\right)^{A/C_p^0} \exp\left(\frac{B}{A}(T - T_0) + \frac{C}{2A}(T^2 - T_0^2) + \frac{D}{3A}(T^3 - T_0^3)\right). \qquad (4)$$

In terms of this new variable, the change of entropy of the gas can be written as:

$$ds = c_p \frac{dT}{T} - R\frac{dP}{P} = c_p^0 \frac{d\tau}{\tau} - R\frac{dP}{P} \qquad (5)$$

and this allows us to define the potential temperature as:

$$\theta = \tau \left(\frac{P}{P_0}\right)^{-\kappa^0} \qquad (6)$$

where, $\kappa^0 = R^*/C_p^0$. In this study, we have used $P_0 = 0.987$ bar and $T_0 = 134$ K as reference values. Under our assumption of no ortho-para exchange, this definition of potential temperature would agree with the definition given by Gierasch et al. (2004) if the reference temperature $T_0$ were high enough to verify $C_p^0 \approx C_p(T \to \infty)$. Our chosen value of $T_0$ is not high enough, and thus our potential temperature differs from theirs, but it is still constant in isentropic surfaces.

Figure 2 shows potential temperature maps between 68° latitude and the poles, extending from 1 mbar to 1 bar pressure for December 2008 (A, D), June 2013 (B) and October 2006 (C). These maps show a similar behaviour of the potential temperature at both Polar Regions, where isentropic surfaces essentially follow isobaric surfaces, with a rapid large vertical change of the potential temperature for pressures above 60 mbar level. The potential temperature reaches values of ~ 1485 K and ~ 1635 K in the upper stratosphere around 1 mbar in the north and south Polar Regions respectively, and decreases rapidly to around 130 K at 1 bar. No seasonal variations of the potential temperature are observed in the south polar regions, while larger values of potential temperature between 2 mbar and 1 mbar are observed in the north polar regions, due to the large temperature difference at these pressures.

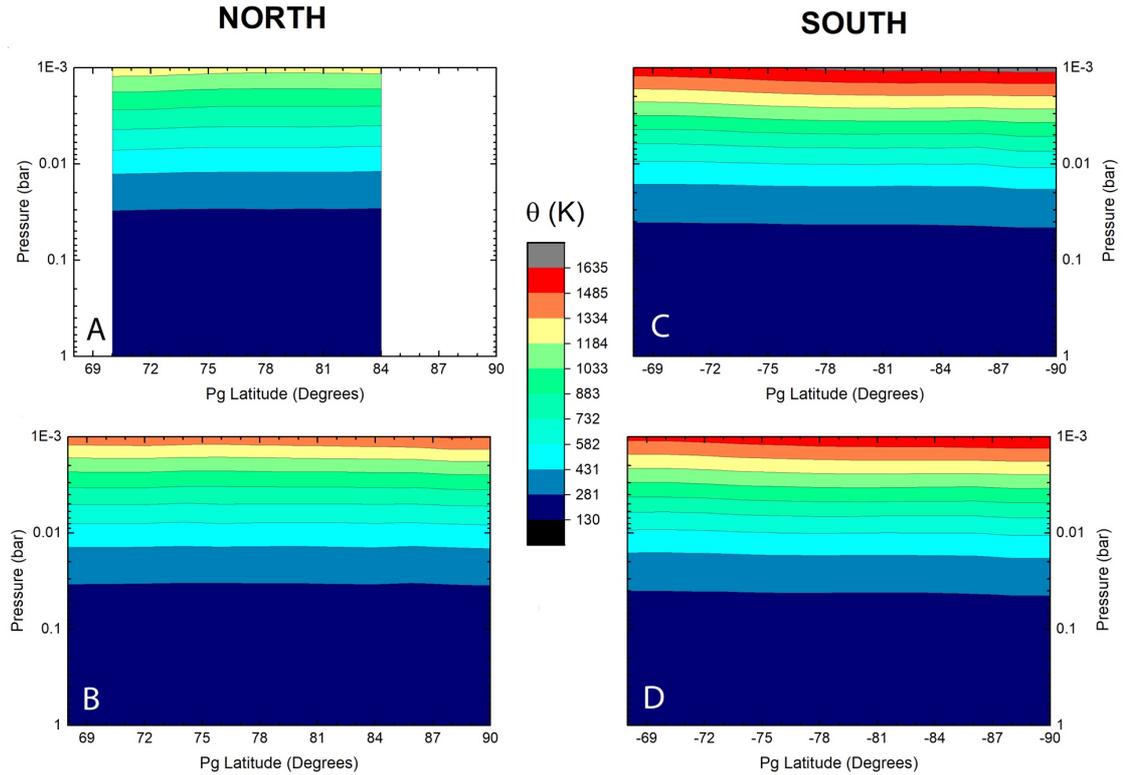

Figure 2. Zonally averaged potential temperature maps from 68° to the poles from December 2008 at both polar regions (A, D), June 2013 at the north polar region (B) and from October 2006 at south polar region (C).

The vertical gradient of the potential temperature is an important quantity that gives us information of the static stability of the atmosphere, essential to understand the nature of different features in the cloud field. Upper and middle rows in Figure 3 show maps of the vertical gradients of the potential temperature between 1 bar and 1 mbar of both polar regions for the epochs under study. These results show that in the stratosphere d$\theta$/d$P$ is negative at all the studied epochs, with d$\theta$/d$P$ growing in magnitude with altitude, showing that the stratosphere is statically stable, and therefore stratified, the upper being more statically stable than the lower stratosphere.

As the vertical gradient of the potential temperature varies strongly with height, the colour scale of these maps does not allow us to observe whether the troposphere and the lower stratosphere are statically stable or neutrally stable (d$\theta$/d$P=0$). In order to analyse the behaviour at lower altitudes, we have plotted the potential temperature ($\theta$) and its vertical gradient (d$\theta$/d$P$) as a function of pressure for the latitude of the core of the Hexagon (i.e. 78° N) and the zonal jet in the south at ~ 74° S in Figure 3c and f. The results show that d$\theta$/d$P$ is negative above ~600 mbar pressure level, indicating that at all epochs, and at both polar regions, the stratosphere and the upper troposphere are statically stable, essential for dynamical instabilities of barotropic or baroclinic character to exist. Between 1 bar and 600 mbar the atmosphere appears to be neutrally stable (see Figure 3f). Finally, Figure 3c does not show any significant seasonal variability and no differences are observed in the vertical gradient of the potential temperature between the Hexagon and the latitude of the zonal jet in the south.

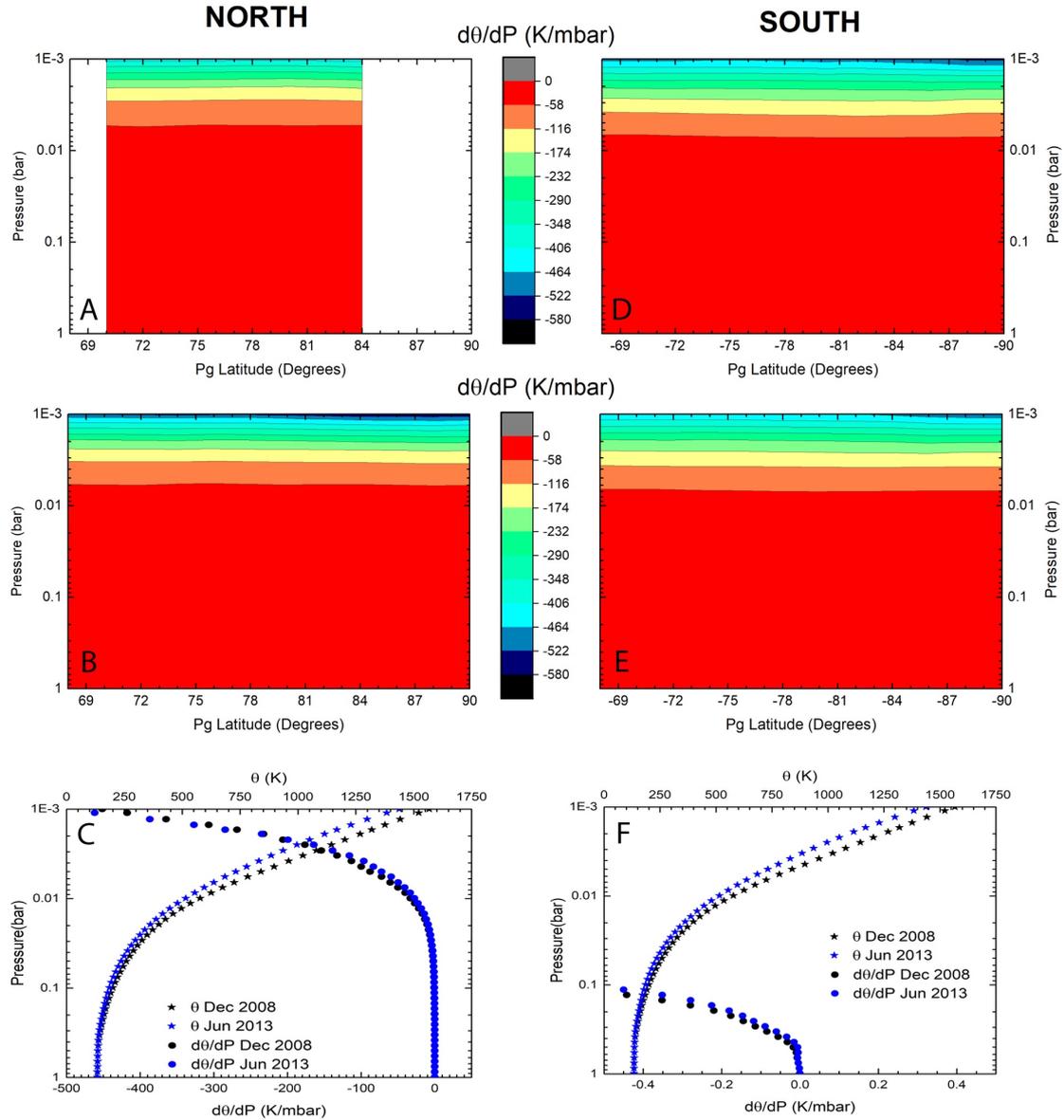

Figure 3. Vertical gradients of the zonally averaged potential temperature maps from 68° to the poles from December 2008 at both polar regions (A, E), June 2013 at the north polar region (B) and from October 2006 at south polar region (D). Panel C shows the extended potential temperature $\theta$ (stars) and its vertical gradient d$\theta$/dP (dots) as a function of pressure for June 2013 (blue) at 78° N planetographic and for December 2008 (black) at 74° S, showing the stratification of the atmosphere. Panel F is a zoom of (C) in order to show the deviation of d$\theta$/dP (dots) from zero.

## 2.3. Quasi-geostrophic potential vorticity.

As mentioned above, generally, in adiabatic and frictionless flows, Ertel potential vorticity is conserved on isentropic surfaces, implying that changes in the absolute vorticity are related to changes in the vertical gradient of these isentropic surfaces. Ertel potential vorticity is conserved in adiabatic motion provided that the potential temperature is a function of density and pressure alone. This is the case in a frozen ortho-para distribution, where specific heat is fixed once temperature and pressure (and thus density) are specified. More complicated dynamics arises if ortho-para exchange is allowed (*Gierasch et al. 2004*).

In rapid rotating fluids, such as Saturn's polar atmosphere, where the Rossby number ($R_0 = U/fL$, where U is the horizontal velocity, $f = 2\Omega\sin\phi$ is the Coriolis parameter and L is the horizontal length-scale) is smaller than one, large-scale winds are in geostrophic balance and therefore, the quasi-geostrophic analysis provides a better approximation of the potential vorticity than that of the Ertel potential vorticity. In this study, we consider a geostrophic base atmosphere, with properties that depend only on height (in our case vertical profiles computed as the area-weighted latitudinal averages at constant pressure, denoted by "$< >$" in eq. 7 and eq. 8), and small geostrophic perturbations on this base atmosphere. To take into account the dependence of specific heat with temperature, we follow again *Gierasch et al.* (*2004*) and *Read et al.* (*2009*) and define the quasi-geostrophic potential vorticity as:

$$q_g = (f + \xi_g) - f \frac{\partial}{\partial P}\left(\frac{P}{s(P)} \frac{T'}{<T>}\right) \qquad (7)$$

where $\xi_g$ is the relative vorticity associated to geostrophic motion, $T' = T - <T>$ and the static stability parameter $s(P)$ is defined as

$$s(P) = -\frac{C_P^0}{<C_p>} \frac{\partial <\ln\theta>}{\partial \ln P} \qquad (8)$$

where $C_P^0$ is the specific heat capacity at 1 bar level, where we assume hydrogen-helium mixture in equilibrium at high temperatures (*Read et al., 2009*).

In order to estimate the error introduced in the QGPV due to errors on the zonal wind profiles (with a mean standard deviation of 7 m s$^{-1}$, *Antuñano et al., 2015*), and the temperature profiles (around 2 K at the stratosphere above $\tilde{\tau} = 140$ K and 0.5 K at the troposphere, *Fletcher et al., 2015*), we have added random perturbations of a magnitude equal to the error to the zonal wind and temperature profiles, respectively, and analysed the variations in QGPV caused by these perturbations. Results are discussed in section 3.1.

**2.4. Thermal wind and relative vorticity**

Under hydrostatic and geostrophic balance, the vertical structure of the horizontal wind field and the horizontal temperature gradient are related by the thermal wind equation:

$$\frac{\partial u}{\partial \ln(P)} = \frac{R^*}{f}\left(\frac{\partial T}{\partial y}\right)_P \qquad (9)$$

where *y* is the meridional coordinate. In this study, we have assumed that the cloud level at which winds have been measured corresponds to a pressure level of 500 mbar (*Sanz-Requena et al., 2017*), which corresponds to a zonally averaged isentropic surface of $\theta \sim 137$ K. In order to evaluate the zonal wind profile with the same latitudinal resolution as the thermal structure and obtain smooth distributions, we have binned the zonal wind profile in 0.5° latitude and then, we have used data separated by 2° latitude. We have then used the wind values at this level as a boundary condition for integration of the thermal wind equation.

Figure 4 and Figure 5 shows the vertical wind shear (a, d), winds at different heights derived from the thermal wind equation (b, e) and the relative vorticity (c, f) at different pressure levels between 1 mbar and 500 mbar for June 2013 and December 2008 in the north polar regions (Figure 4) and for October 2006 and December 2008 in the south polar region (Figure 5). Comparing the vertical wind shear of both polar regions at all epochs, we find that in both regions the vertical wind shear in the troposphere is smaller than in the stratosphere, ranging between $\left|\frac{du}{dln(P)}\right| = \pm 4$ m s$^{-1}$ in the troposphere ($du/dz \sim \pm 0.1$ m s$^{-1}$/km, assuming a mean height scale of $H \sim 40$ km) and reaching

maximum values of about $\left|\frac{du}{d\ln(P)}\right| = 10$ m s$^{-1}$ in the north and ($du/dz \sim -0.25$ m s$^{-1}$/km) and $\left|\frac{du}{d\ln(P)}\right| = 20$ m s$^{-1}$ ($du/dz \sim -0.5$ m s$^{-1}$/km) in the south, respectively. On the other hand, seasonal variations of the vertical wind shear are mainly significant at around 10 mbar in the south polar region near 87° S, where it is ~1.5 times larger in October 2006 (southern summer) than in December 2008 (southern late summer), and in the north polar region, both in the stratosphere and upper troposphere, indicative of a tendency towards a more baroclinic atmosphere during both northern and southern summer. In addition, Figure 4b,e and Figure 5b,e indicate that, overall, westward jets present positive vertical wind shear while eastward jets display negative vertical wind shear with height. The results of the vertical wind-shear are in agreement with previous studies (*Read et al., 2009; Fletcher et al., 2016*), however, they are around 4 times smaller than those used by *Morales-Juberías et al.* (*2015*) in their numerical study of the Hexagon jet.

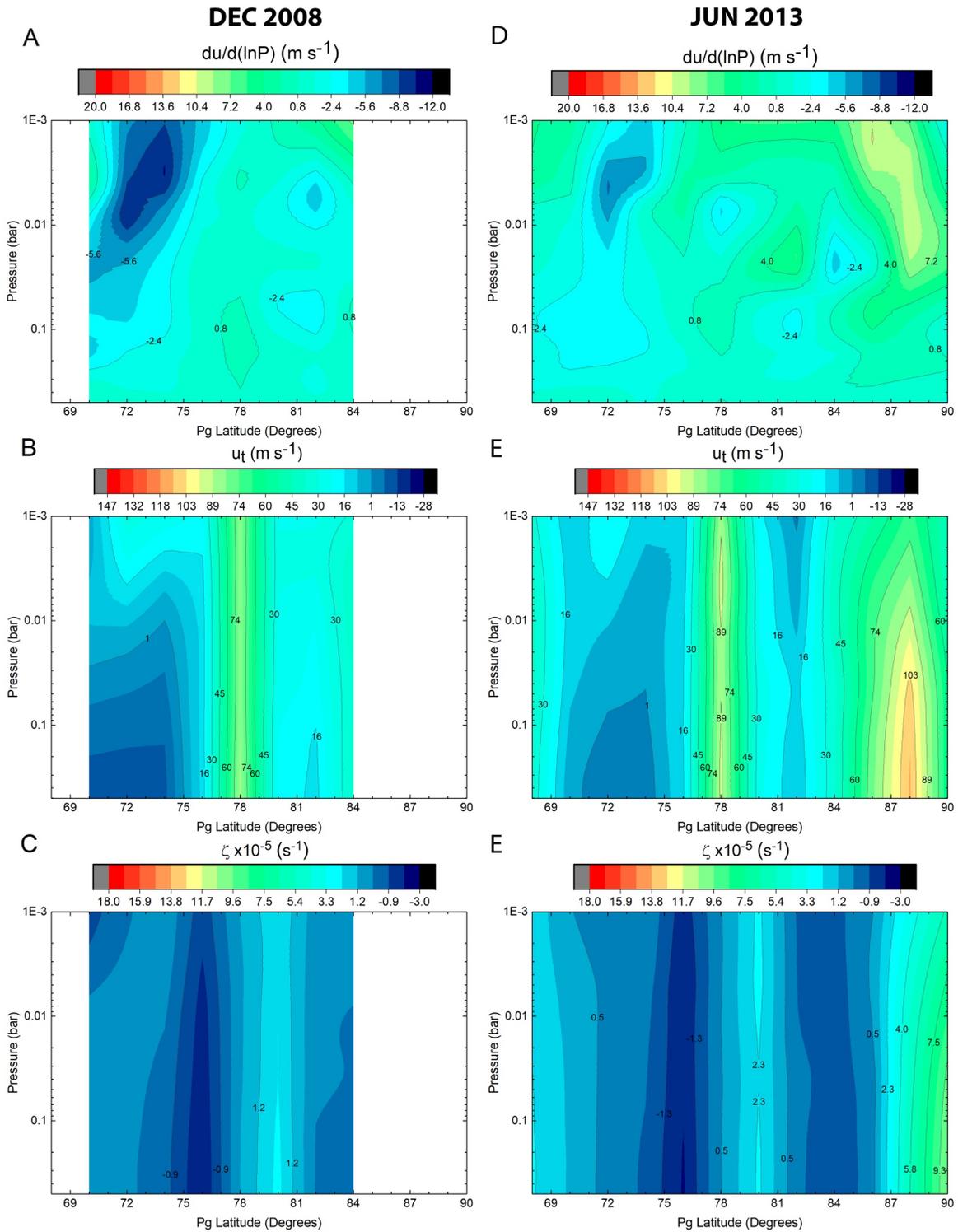

Figure 4. Pressure-latitude wind-shear map for December 2008 (A) and June 2013 (D), integrated zonal wind velocity map for December 2008 (B) and June 2013 (E) and relative vorticity maps from December 2008 (C) and June 2013 (F) from 68° and the pole and between 1 mbar and 500 mbar pressure levels for the north polar region.

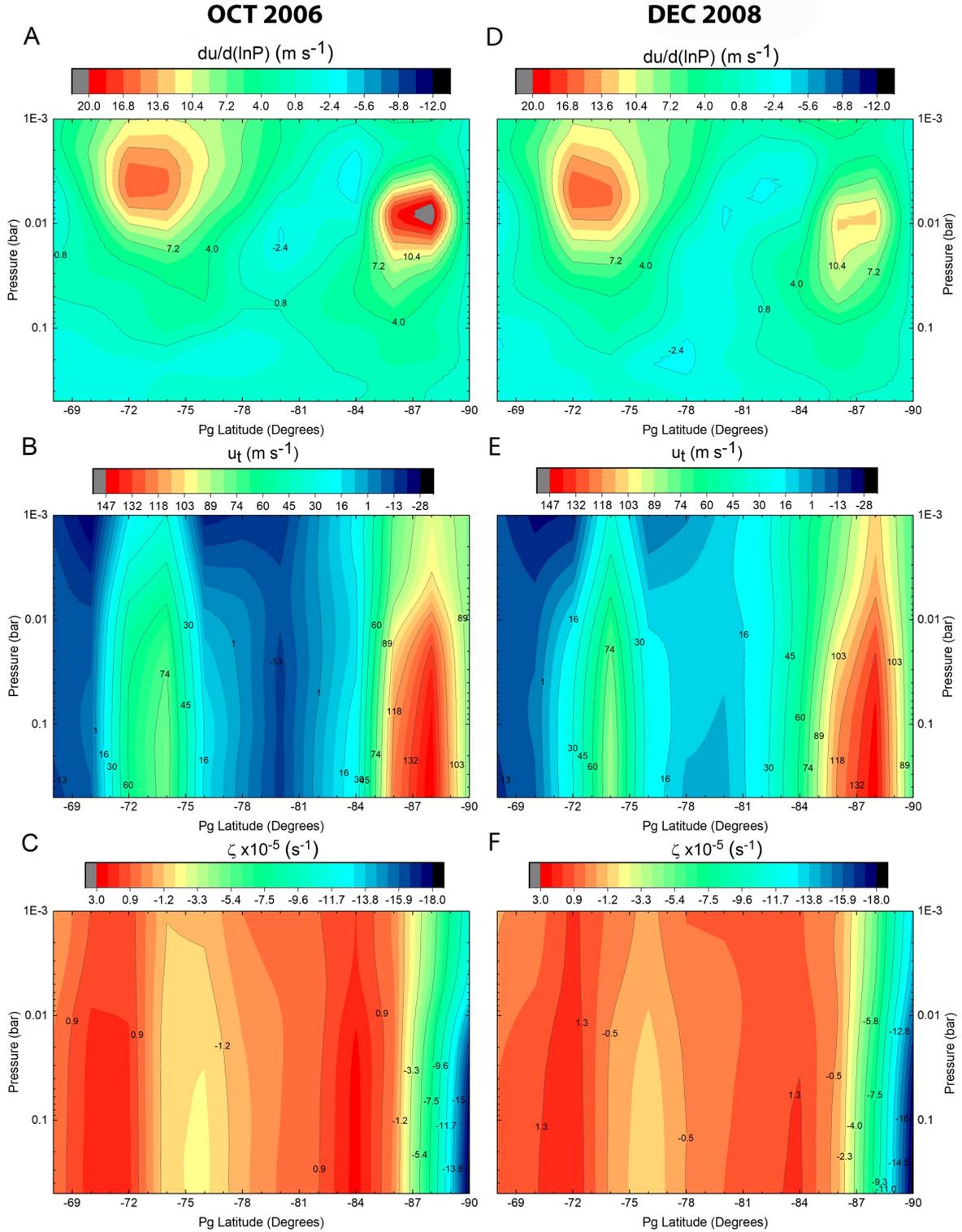

Figure 5. Same as in Figure 4, but for October 2006 (left column) and December 2008 (right column) 68° S and the pole and between 1 mbar and 500 mbar pressure levels for the south polar region.

Relative vorticity maps shown in Figure 4c,f and Figure 5c,f have been computed at each pressure level using the following expression, (*Sánchez-Lavega et al., 2006*)

$$\xi(\lambda, \phi) = -\frac{1}{R(\phi)} \frac{\partial u}{\partial \phi} + \frac{u}{R(\phi)} \tan\phi \qquad (10)$$

where the term related to the meridional wind is neglected due to the large meridional wind shear of the zonal flow, $u$ is the wind field presented in Figs 4 and 5, $\lambda$ is longitude, $\phi$ is latitude and $R(\phi)$ is the radius of the planet at latitude $\phi$, given by $R(\phi) = R_e R_p (R_e^2 \sin^2\phi + R_p^2 \cos^2\phi)^{-1/2}$. The last term on the right of eq. 10, usually ignored at lower latitudes, becomes important at the core of the polar jet where it reaches approximately the same order of magnitude of the first term on the right (*Antuñano et al., 2015; Sayanagi et al., 2017*). The tangent in eq. 10 is singular at the pole. We deal with this singularity using the fact that the winds decrease approximately linearly with colatitude at latitudes higher than 88° north and south, and approximating the last term in eq. 10, using the small-angle approximation:

$$\frac{u}{R}\tan(\phi) = \frac{1}{R}\frac{a(\pi/2-\phi)}{\tan(\pi/2-\phi)} \sim -\frac{a}{R} \qquad (11)$$

where $a$ is the slope of the zonal velocity profile between 90° and 88.5° north and south (e.g., $a = 5345$ m s$^{-1}$ rad$^{-1}$ and $a = 6113$ m s$^{-1}$ rad$^{-1}$ at cloud level in the north and south respectively).

The results show that, due to the binning of the zonal wind profile that leads to slower zonal winds at the jet peaks, we obtain a lower relative vorticity than that found in *Antuñano et al.* (*2015*) and *Sayanagi et al.* (*2017*). Introducing the term related to longitudinal gradients of the meridional winds in eq. 10 does not change these results. In addition, no significant seasonal variations of the relative vorticity are observed.

3. **Normalised zonally averaged potential vorticity**
    3.1. **Quasi-geostrophic Potential Vorticity**

Zonally averaged QGPV maps for October 2006 (south), December 2008 (north and south) and June 2013 (north) are shown in Figure 6. This figure shows that QGPV is positive at all pressure levels in the north polar region, while it is negative in the south polar region, showing the dominance of the Coriolis parameter, $f$. This is everywhere an order of magnitude larger than the relative vorticity except in a narrow region near the pole, where both quantities have the same order of magnitude. At both poles, and at all altitudes, the absolute value of QGPV grows with latitude, following the tendency of the rest of the planet (Read et al 2009). QGPV absolute values range between 2.3 x 10$^{-4}$ s$^{-1}$ and 4.5 x 10$^{-4}$ s$^{-1}$ in the North Polar Region and between -2.8 x 10$^{-4}$ s$^{-1}$ and -5.2 x 10$^{-4}$ s$^{-1}$ in the south, reaching its maximum value near the poles at around 100 mbar pressure in the North Polar Region and at around 300 mbar in the south. These results present the highest standard deviation of the estimated error at the troposphere, between 500 mbar and 200 mbar at both polar regions, displaying maximum standard deviation values of ~ 0.4 x 10$^{-4}$ s$^{-1}$ in the south and ~0.8 x 10$^{-4}$ s$^{-1}$ in the north polar region. At higher altitudes, the standard deviation decreases, reaching maximum values of ~0.1x 10$^{-4}$ s$^{-1}$ and ~ 0. 2x 10$^{-4}$ s$^{-1}$ at ~100 mbar in the south and in the north, respectively. In the stratosphere, the average standard deviation of the estimated error of the quasi-geostrophic potential vorticity values are 0.04 x 10$^{-4}$ s$^{-1}$ at the south and 0.07 x 10$^{-4}$ s$^{-1}$ at the north.

As a reference, the absolute QGPV values obtained in this study can be compared to the potential vorticity of the North polar vortex on Earth located at ~450 K isentropic surface (*Nash et al., 1996*), equivalent ~10 mbar altitude pressure level on Saturn (see Figure 2). We find that the QGPV of Saturn's polar regions at the same altitude is around 20 times larger than the potential vorticity of the North polar vortex on Earth.

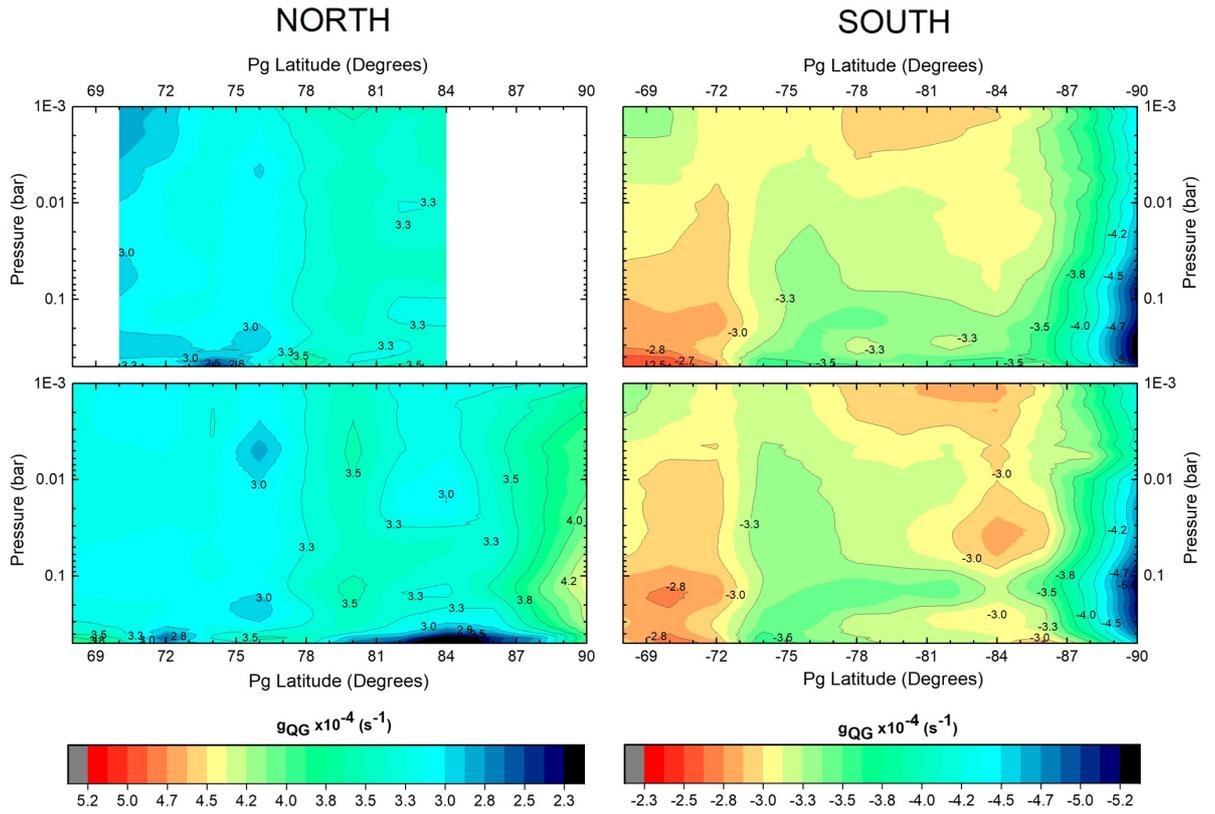

Figure 6. Zonally averaged quasi-geostrophic potential vorticity maps from 68° to the pole for June 2013 (bottom left) and December 2008 (top left) at the North Polar Region and for December 2008 (top right) and October 2006 (bottom right) at the South Polar Region.

Meridional profiles of the quasi-geostrophic potential vorticity of the north and south polar regions at 255 mbar, 86 mbar, 10 mbar and 2 mbar are given in Figure 7. As is apparent from this figure, the quasi-geostrophic potential vorticity displays a step-like behaviour related to the eastward jets found in the polar regions. This step-like behaviour is observed at all the altitudes shown in Figure 7. However, the steps seem to be more prominent in the troposphere than in the stratosphere, potentially due to the decreasing velocity of the zonal eastward jets with altitude (see Figure 4 and Figure 5), in agreement with the step-like behaviour reported by Read et al. (2009). Furthermore, these steps are observed to be narrower in the north polar region compared to the south polar region, due to the narrower jets found in this region compared to the south (*Antuñano et al., 2015*). Finally, the most prominent steps are found at around 84-86° N and 84° S, associated to the strong polar jets, where the potential vorticity increases rapidly poleward of these latitudes.

The QGPV maps and profiles presented in this study differ from those in Read et al. (2009) mainly in the stratosphere, where this study presents QGPV values around 15% larger in magnitude at the southern stratosphere (*2009*) and around 15% smaller in the northern stratosphere. In the northern troposphere (around 250-290 mbar), this study shows values ~10% larger than those in Read et al. (2009), and in the southern troposphere both studies give similar results, with differences smaller than 5%. These differences might be due to the higher resolution wind and temperature profiles used in this study and the lower smoothing performed here.

In the simulations of the Hexagonal jet of Morales-Juberías et al. (2011), values of PV at 200 mbar are in the range of $4.4$-$5.6 \times 10^{-4}$ s$^{-1}$ once they are normalized as in this study (dividing their PV by $g \langle d\theta/dP \rangle$, where g is the gravity and "$\langle \ \rangle$" indicates the area-weighted latitudinal average at a constant pressure). These values are around 1.5 times larger than the results we report here.

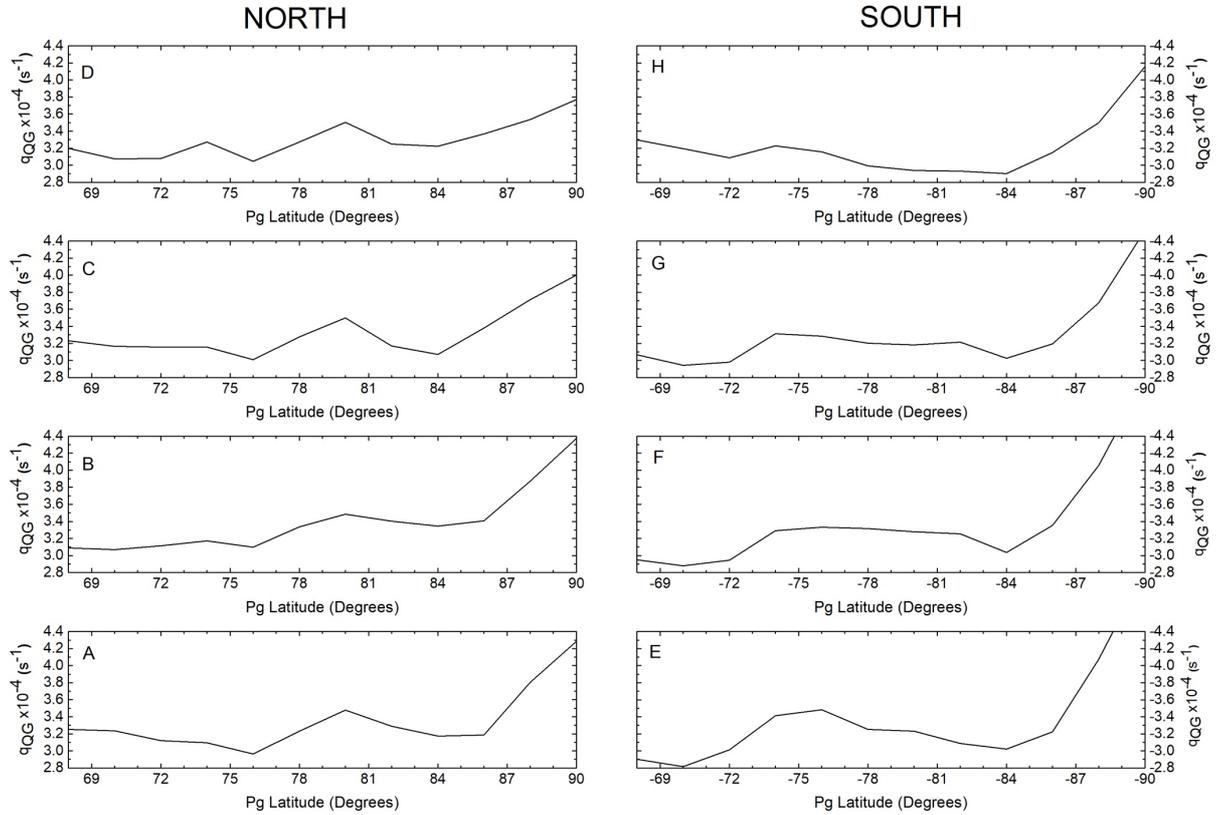

Figure 7. Meridional profiles of quasi-geostrophic potential vorticity maps of (left) the north polar region and (right) the south polar region at four different altitudes (A, E) 2 mbar, (B, F) 10 mbar, (C, G) 86 mbar and (D, H) 255 mbar, showing the staircase of the potential vorticity, mainly at the north polar region.

Analysing the seasonal variations in both polar regions between northern late winter (December 2008) and northern early spring (June 2013), as well as between southern summer (October 2006) and southern late summer (Dec 2008), we find that the small variations of the QGPV observed in both polar regions are within our estimated error and thus, are not significant (see Figure 6). In the case of the North Polar Region, this is potentially due to the dominance of the Coriolis parameter at the latitudes available for seasonal comparison and we might expect seasonal variations at higher latitudes (i.e. poleward of 88° N), where the relative vorticity is of the order of magnitude of the Coriolis parameter and therefore, the seasonal temperature differences start to be significant. In the South Polar Region, where we have access to a comparison all the way to the pole, we have not detected any seasonal differences, probably due to the short temporal interval between the two datasets (two years in the south compared to 5 years in the north).

Meridional gradients of the zonally averaged QGPV maps for the northern spring (June 2013), northern late winter (December 2008), southern summer (October 2006) and southern late summer (December 2008) are shown in Figure 8. These maps show that $dq_g/dy$ changes sign clearly at the southern flank of the polar jets at ~84° north and south, at both flanks of the Hexagon jet and of the eastward jet at 73.9° S, and for some specific pressure levels at some latitudes not related to the zonal wind boundaries. This indicates that the Charney-Stern necessary (but not sufficient) criterion for baroclinic instabilities to grow is mostly satisfied, in agreement with the results from *Read et al.* (*2009*) and *Fletcher et al.* (*2016*). These results do not show any differences between the Hexagon jet and the zonal jet at 73.9° S that could explain the presence of the Hexagon in the northern hemisphere and its absence in the south polar region. Finally, as we have mentioned previously, no significant seasonal variation is observed.

Although these results show the tendency of the meridional gradients of the potential vorticity to change sign at certain latitudes, the precise values must be taken with caution as this magnitude depends on differentiated velocity and temperature fields and therefore, the error propagation of the measurement uncertainties in the horizontal wind field and the retrieved temperature profile become important.

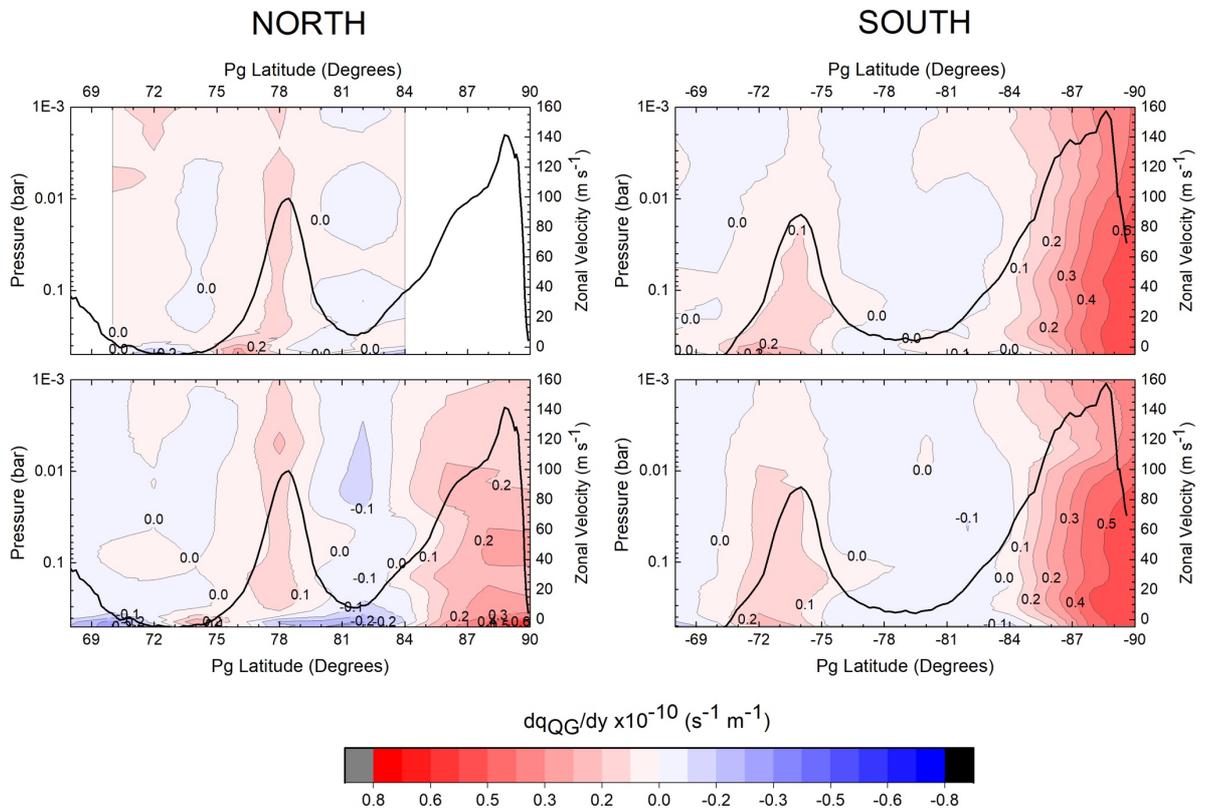

Figure 8. Meridional gradients of the zonally averaged quasi-geostrophic potential vorticity maps between 68° and the pole for December 2008 (top left) and June 2013 (bottom left) in the North Polar Region and December 2008 (top right) and October 2006 (bottom right) in the south (bottom), showing the change in sign at the flanks of the Hexagon, eastward jet at 74.2° S and the equatorward flank of the polar jets. The solid black lines represent the zonal wind profiles of both polar regions at cloud level.

In order to analyse the contribution of the variations in the stratification of the atmosphere (i.e. the contribution of a baroclinic term) to the meridional gradient of the quasi-geostrophic potential vorticity, we compare the latter with the meridional gradient of the absolute vorticity of a pure barotropic atmosphere $(d(f + \xi_g)/dy)$. Figure 9 shows the meridional gradients of the absolute vorticity, $d(f + \xi_g)/dy$, for the four studied epochs. Comparing Figure 8 and Figure 9, we observe that the barotropic term accounts for the main tendencies of the meridional gradients of the quasi-geostrophic potential vorticity, leading to potential growth of dynamical instabilities at the flanks of the Hexagon, its counterpart in the south and at the southern flank of the polar jets at all the studied pressures and epochs, as it clearly changes sign in these regions. When we add the baroclinic term, we observe that $dq/dy$ changes sign overall at the same regions, but this time the changes do not strictly follow the latitudinal boundaries of the jets, varying with pressure and resulting in a larger number of latitudes where dynamical instabilities (either barotropic or baroclinic) could develop. These results do not show any differences between the Hexagon jet and the jet at 73.9° S that could explain the formation of a Hexagon only in the north polar region.

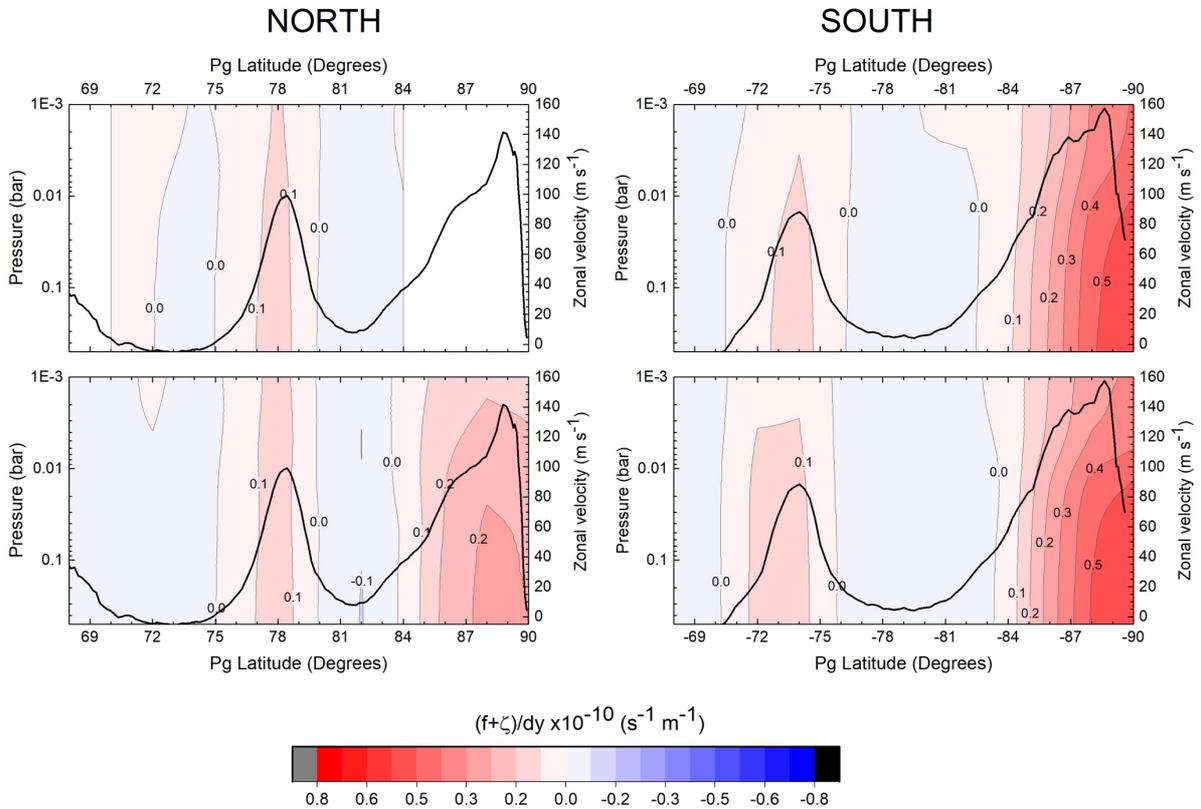

Figure 9. Meridional gradients of the absolute vorticity, $d(f + \xi_g)/dy$, for December 2008 (top left and right), June 2013 (bottom left), October 2006 (bottom right), showing the change in sign at the flanks of the Hexagon, the eastward jet at 74.2° S and the equatorward flank of both polar jets, at all the studied epochs. The solid black lines represent the zonal wind profiles of both polar regions at cloud level.

## 4. Discussion

In this study, we compute quasi-geostrophic potential vorticity (QGPV) maps of Saturn's north and south polar regions between 68° and 90° planetographic latitude for different pressures between the upper troposphere (500 mbar) and upper stratosphere (1 mbar) for June 2013 (north), December 2008 (north and south) and October 2006 (south). For this purpose, we have used the high-resolution zonal wind and temperature profiles derived by the Cassini spacecraft. Additionally, we study the stability of the zonal jets in both polar regions, by analysing the meridional gradients of the QGPV. We draw the following conclusions:

- **QGPV**: The overall behaviour of the QGPV is very similar in both polar regions, where the vorticity dynamics is clearly dominated by the Coriolis parameter. This is an order of magnitude larger than the relative vorticity everywhere except at the pole, leading to positive QGPV values in the northern hemisphere and negative in the southern region. QGPV reaches maximum values of ~ 4.5 x $10^{-4}$ $s^{-1}$ in the North Polar Region and ~ -5.2 x $10^{-4}$ $s^{-1}$ in the south near the poles in the upper troposphere, due to the strong cyclonic vorticity of the polar vortices at cloud level, which weakens with altitude due to the negative wind-shear. Overall, the absolute values of QGPV grow with latitude at all pressure levels and epochs, reaching their maximum value at the poles, following the tendency of the rest of the planet.

- **Potential vorticity gradients:** We present maps of $dq_g/dy$ to diagnose stability conditions of Saturn's north and south polar regions between 68° and 90° latitude for the same dates, by computing and analysing meridional gradient maps of the normalised QGPV. The results show that at the studied epoch dynamical instability (regardless of barotropic or baroclinic character) could develop on the flanks of both the Hexagon jet at 78° N and the eastward jet at 73.9° S planetographic and on the southern flanks of both north and south polar jets at ~ 89°. We observe that these instabilities could also develop in other regions and some specific pressures. A comparison of meridional gradients of the total QGPV and its barotropic component shows that the barotropic term is responsible for the locations where these kind instabilities could develop. We do not observe any significant differences in the stability of the northern hexagon jet and its southern counterpart at 73.9° S that could explain the presence of the hexagon in one hemisphere but not the other.

- **Temporal variability:** We present a study of the temporal variability of the QGPV. We do not detect any significant variability over an interval of nearly 7 Earth years or ~0.25 of a Saturnian year, and thus any possible seasonal variation of the QGPV lies within our estimated errors. These results are in agreement with the expectations because, although the temperature field has been observed to change strongly with seasons, its meridional gradient did not change significantly, resulting in small temporal variations of the vertical wind-shear and therefore, small temporal variation of the relative vorticity, which is, together with the Coriolis parameter, the dominant term in the potential vorticity.

So far, diverse numerical simulations (Morales-Juberías et al., 2011; Morales-Juberías et al., 2015, Rostami et al., 2017), analytical approaches (Barbosa-Aguiar et al., 2010; Sánchez-Lavega et al., 2014; Antuñano et al., 2018), as well as laboratory experiments (Barbosa-Aguiar et al., 2010) have been performed in order to study the nature and origin of the Hexagonal jet. However, this is still an open question. The similarities of the potential vorticities and vorticity gradients at both polar regions and at all the studied epochs shown in this study do not give us any further indication of why a Hexagon wave is present in the north polar region and not in the south. The study of the three-dimensional potential vorticity may give us more information regarding this intriguing issue. However, the Cassini CIRS data do not have sufficient signal to noise or coverage to provide longitudinally-resolved maps of the hexagon that are coincident with the wind maps and therefore, new spacecraft/telescope data will be needed. It is possible that the James Webb Space Telescope could provide both temperatures and winds in this region in the coming decade, with enough spatial resolution to compute 3D potential vorticity maps. Additionally, new datasets of simultaneous wind and temperature profiles will allow us to further our study in order to improve our knowledge on the relationship of the seasonal thermal effects and the polar atmospheric dynamics.

**Acknowledgements**
The data for this paper are available at Antuñano et al. (2015), García-Melendo et al. (2011), Sánchez-Lavega et al. (2006) and Fletcher et al. (2015, 2016). ASL and TdRG acknowledge support by the Spanish project AYA2015-65041-P (MINECO/FEDER, UE) and Grupos Gobierno Vasco IT-765-13. AA and LNF were supported by a European Research Council Consolidator Grant under the European Union's Horizon 2020 research and innovation programme, grant agreement number 723890, at the University of Leicester. LNF was also supported by a Royal Society Research Fellowship. PLR acknowledges support from the UK Science and Technology Facilities Council under grants ST/K00106X/1 and ST/N00082X/1.